\documentclass[%
reprint,
superscriptaddress,
nofootinbib,
 amsmath,amssymb,
 aps, prb,
]{revtex4-2}

\usepackage{siunitx}
\usepackage{graphicx}
\usepackage{layouts}
\usepackage{multirow}
\usepackage{footmisc}
\usepackage{overpic}
\usepackage[normalem]{ulem}

\def\<#1>{\mathinner{\langle#1\rangle}}

\usepackage[T1]{fontenc}

\usepackage{booktabs}

\begin{document}
\vspace{-0.5cm}

\title{Coherence limit due to hyperfine interaction with\\nuclei in the barrier material of Si spin qubits}

\author{Lukas Cvitkovich}
\email[]{cvitkovich@iue.tuwien.ac.at}
\affiliation{Institute for Microelectronics, Technische Universität Wien}



\author{Peter Stano}
\affiliation{RIKEN Center for Emergent Matter Science (CEMS), Wako, Japan}
\affiliation{Institute of Physics, Slovak Academy of Sciences, Bratislava, Slovakia}

\author{Christoph Wilhelmer}
\affiliation{Institute for Microelectronics, Technische Universität Wien}

\author{Dominic Waldhör}
\affiliation{Institute for Microelectronics, Technische Universität Wien}

\author{Daniel Loss}
\affiliation{RIKEN Center for Emergent Matter Science (CEMS), Wako, Japan}
\affiliation{Department of Physics, University of Basel, Basel, Switzerland}

\author{Yann-Michel Niquet}
\affiliation{Univ. Grenoble Alpes, CEA, IRIG-MEM-L Sim, F-38000, Grenoble, France}

\author{Tibor Grasser}
\affiliation{Institute for Microelectronics, Technische Universität Wien}


\begin{abstract}

On the quest to understand and reduce environmental noise in Si spin qubits, hyperfine interactions between electron and nuclear spins impose a major challenge. 
Silicon is a promising host material because one can enhance the spin coherence time by removing spinful $^{29}$Si isotopes. 
As more experiments rely on isotopic purification of Si, the role of other spinful atoms in the device should be clarified. This is not a straightforward task, as the hyperfine interactions with atoms in the barrier layers are poorly understood.
We utilize density functional theory to determine the hyperfine tensors of both 
Si and Ge in a crystalline epitaxial Si/SiGe quantum well as well as 
Si and O atoms in an amorphous Si/SiO$_2$ (MOS) interface structure.
Based on these results, we estimate the dephasing time $T_2^*$ due to magnetic noise from the spin bath 
and show that the coherence is limited by interactions with non-Si barrier atoms to a few µs in Si/SiGe (for non-purified Ge) and about 100\,µs in Si-MOS. Expressing these numbers alternatively, in Si/SiGe the interactions with Ge dominate below 1000\,ppm of $^{29}$Si content, and, due to low natural concentration of the spinful oxygen isotopes, the interactions with oxygen in Si-MOS become significant only below 1\,ppm of $^{29}$Si content.

\end{abstract}
\maketitle


\section{Introduction}
Spin qubits implemented in gated semiconductor quantum dots~\cite{burkard_semiconductor_2023} are considered promising candidates for the physical implementation of quantum information processing since the proposal by Loss and DiVincenzo~\cite{Loss1998}.
Coherent control of such a spin qubit in a physical device
is impeded by interactions with the environment. Although the relative importance of various decoherence mechanisms is still under debate~\cite{Yoneda2023, Struck2020},
a major source of decoherence is the hyperfine interaction with the nuclear spins of the host material~\cite{khaetskii_electron_2002,merkulov_electron_2002}.
In this regard, silicon is an outstanding platform for implementing spin qubits due to the naturally low abundance of $^{29}$Si, its only stable spinful isotope. 
Removing $^{29}$Si by isotopic purification~\cite{itoh_isotope_2014} improves several key spin-qubit metrics~\cite{stano_review_2022} including coherence, which can be pushed even up to a second for donor-bound electrons in bulk silicon~\cite{Tyryshkin2012}.
The second reason for the auspicious status of silicon is the compatibility with classical semiconductor device fabrication, the metal–oxide–semiconductor (MOS) technology~\cite{Angus2007}.
Hence, the development of Si-MOS spin qubits is a direction currently pursued, with encouraging achievements such as high coherence~\cite{veldhorst_addressable_2014, hansen_implementation_2022, elsayed_low_2022}, high-fidelity single and two-qubit gates~\cite{tanttu_consistency_2023}, fast and high-fidelity single-shot readout~\cite{niegemann_parity_2022, oakes_fast_2023}, high quality factors~\cite{leon_bell-state_2021, camenzind_spin_2022}, or the operation above 1 K~\cite{huang_high-fidelity_2023}. As an alternative, Si/SiGe devices rely on a two-dimensional electron gas that is well separated from the heterostructure interface offering a clean system~\cite{esposti_low_2023} and coherent \cite{yoneda_quantum-dot_2017,struck_low-frequency_2020,mills_two-qubit_2022,weinstein_universal_2023} and high-fidelity \cite{takeda_resonantly_2020,blumoff_fast_2022,philips_universal_2022,mills_high-fidelity_2022} spin qubits.

In this paper, we employ density functional theory (DFT) to determine the hyperfine interactions between a conduction band electron in a Si quantum dot and a nuclear spin bath that is composed of atoms in the Si host lattice and the barrier layer. 
While previous \textit{ab-initio} studies focused on donor-bound electrons~\cite{Swift2020} or delocalized electrons in bulk~\cite{ChrisG1993, Philippopoulos2020}, we consider 
realistic interfaces to which a conduction-band electron is confined by an external electric field. This model allows us not only to calculate the hyperfine interaction with $^{29}$Si, which is the dominant source of decoherence in natural Si, but also with spinful isotopes in the barrier, a weaker interaction with the evanescent tail of the electron that only becomes important for highly purified Si.
Besides $^{29}$Si, spinful isotopes appearing in the barrier include $^{73}$Ge in SiGe and $^{17}$O in MOS structures.
Based on the DFT results, we estimate dephasing times $T_2^*$ due to quasi-static noise from nuclei in the barrier.

For Si/SiGe, we come to the conclusion that $^{73}$Ge isotopes limit the coherence time to a few µs already at 3500\,ppm of $^{29}$Si fraction. This value is less than 4.68\%, the natural abundance of $^{29}$Si, but substantially more than 50\,ppm, the lowest currently available fraction of $^{29}$Si in purified silicon.
Considering isotopically purified Ge in addition to purified Si will suppress the nuclear noise further and, according to our model, allows to improve the decoherence time by another order of magnitude.

For Si-MOS, the hyperfine coupling to oxygen atoms in the barrier would overtake the spin-qubit dephasing if the isotopic purification of silicon would be increased by almost two orders of magnitude compared to the lowest currently used value of 50\,ppm. 
While these conclusions depend on external parameters, especially the confining electric potential, we conclude that the spinful non-Si isotopes  will not be a limiting factor of Si-MOS qubits unless the silicon is purified to levels much below 50 ppm.

\section{Methodology}

\begin{figure*}[tbp]
	\centerline{\includegraphics[width=0.7\linewidth]{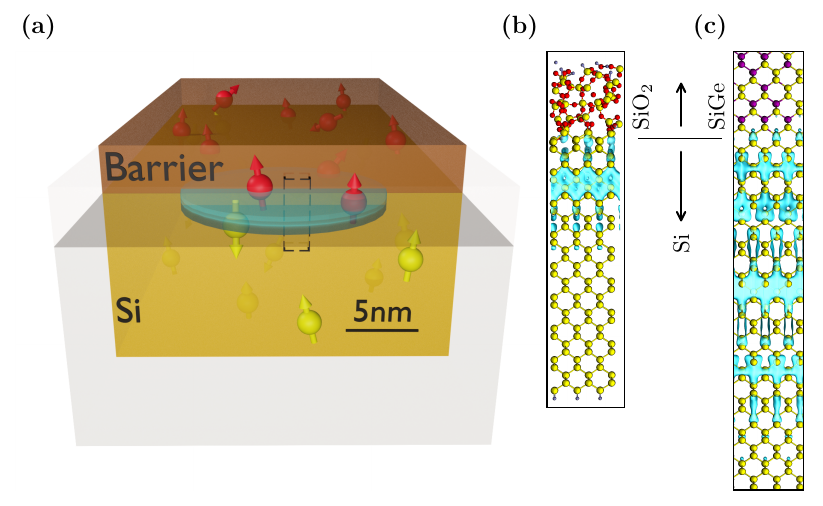}}
    \caption{
    \textbf{(a)}~Schematic device layout of a quantum dot (depicted as a blue ellipsoid) in a Si-based material stack. The dot is formed at the Si/barrier interface by top gates (not shown) which also confine the electron wavefunction in the lateral ($xy$) plane. Nuclear spins in both material layers (schematically represented as red and yellow arrows) act as a source of magnetic noise. This interaction and the resulting decoherence is mostly determined by the electron wavefunction amplitude at atomic nuclei.
    \textbf{(b)}~Exemplary 3D model of an atomistic Si/SiO$_2$ interface structure (side view). We assume an external electric field $50$\,mV/nm in $z$-direction (perpendicular to the Si-barrier interface). The simulation cell measures 1.15$\times$1.15$\times$15$\,$nm$^3$ and contains 780 atoms (Si in yellow, O in red, H for the passivated dangling bonds in blue). 
    The cyan surface displays the real part of the electron wavefunction as obtained from DFT.
    \textbf{(c)}~A part (7.5\,nm) of the Si/Si$_{0.7}$Ge$_{0.3}$ cell. The full cell of the model is 1.65$\times$1.65$\times$25$\,$nm$^3$ and contains 2500 atoms (Si in yellow, Ge in magenta). The plotted wavefunction corresponds to an electric field of $5.8$\,mV/nm. Because of the lower field, the wavefunction extends further into the Si layer. 
    }
    \label{model}
\end{figure*}

\subsection{DFT setup}
We employ density functional theory (DFT) implemented in the CP2K code~\cite{CP2K} using the semi-local GGA functional PBE~\cite{PBE} for all calculations. We relax the geometries of the interface structures and calculate their electronic ground state imposing periodic boundary conditions in all directions. Since obtaining reliable hyperfine couplings within an \textit{ab-initio} framework requires accurate modeling of the spin density in the vicinity of atomic nuclei~\cite{Ghosh2019}, we use all-electron basis sets. Also, the orbitals are expanded in correlation-consistent polarized double-$\zeta$ basis sets for valence and core electrons~\cite{Dunning1989, Peterson2002, woon1995}. Since the same basis sets are not available for Ge, we used the 6-31G double-$\zeta$ Pople basis sets~\cite{Rassolov2001}, which were previously reported to provide accurate hyperfine couplings~\cite{Jakobsen2019, Hermosilla2011}.
The systems are self-consistently relaxed with a force convergence criterion of 0.02 eV/\AA\, per atom.
The calculation of the hyperfine tensors (including relativistic effects) is based on the CP2K implementation of Declerck~\cite{Declerck06}. 

We include a homogeneous electric field perpendicular to the heterostructure interface, using the Berry-phase formalism~\cite{Vanderbilt_solids_1993, Vanderbilt_surface-charge_1993} as presented in~\cite{Souza2002, Umari2002}. 
This implementation corresponds to imposing closed-circuit boundary conditions with a constant applied bias across the simulation cell. The electric fields in the material layers adjust according to the layer thickness and relative permittivity. 
The values for the electric field $F$ quoted throughout this work refer to the field in the Si layer. 
For the Si/SiGe heterostructure, we apply electric fields $F$ within the typical range of 1 to 10\,mV/nm~\cite{McJunkin2022}.
For the Si-MOS heterostructure, the electric field in the 2DEG depends on the doping density of the Si substrate~\cite{Sabnis1979}. 
Assuming a doping density of at most 10$^{18}$/cm$^3$ in the substrate and a carrier density of at most 10$^{12}$/cm$^2$ in the 2DEG, we estimate the field $F$ in the semiconductor~\cite{Dhar2008} to be below 60\,mV/nm.

\subsection{Atomic structure and supercell}
The high computational costs of the DFT framework put restrictions on the size of our simulation cell. Therefore, we do not consider in-plane confinement and enforce periodic boundary conditions on unit cells with lateral size of 1.15 or 1.6 nm along the interfaces. We extrapolate the calculated quantities to realistic in-plane quantum dot sizes by a straightforward rescaling of the electron wavefunction density, with the details given in Sec.~\ref{sec:scaling}. 
The electric field is applied in the perpendicular direction, pushing the conduction-band electron to the interface, see Fig.~\ref{model}.

An interaction between electron spin and nuclear spins in the barrier is mediated through the exponentially decaying tail of the electron wavefunction.
The interaction strength with an individual atomic nucleus is sensitive to the exact details of this decay. It requires a realistic description of the interface at the atomic level, especially for the Si-MOS case, as we explain below.

\subsubsection{Si/SiGe}
Modeling typical experimental setups, we consider a SiGe layer with roughly 30\% Ge content interfaced with a strained Si layer to get the correct band alignment to form a quantum well in the Si part of the simulation cell. By means of the PBE functional, we find a band offset of 200\,meV, slightly overestimating the experimental value of 150\,meV~\cite{Maiti1998}.
For the generation of the structure, we start from a pure [001] Si crystal with 120 monolayers in $z$ direction and $3\times3$ unit cells along $x$ and $y$. A 4.3\,nm thick region (32 monolayers) on one end of the crystal is assigned to form the SiGe barrier.
In this region, SiGe layers are produced by replacing random atoms of the Si lattice by Ge, such that the Ge proportion in each layer equals 27.7\% (5 out of 18 atoms are Ge). The next 12\,nm of pure Si represents the Si quantum well and the final 6\,nm of vacuum serves as a boundary barrier in the model.\footnote{Alternatively, one could consider sandwiched Si layers (barriers at both sides) as a way to confine the electron instead of using the electric field, and smooth transition between the Si and SiGe parts instead of an abrupt interface. We leave these extensions for the future.}
Dangling bonds at the material/vacuum interfaces are passivated with hydrogen.
The whole cell is relaxed under 1\% of tensile in-plane strain.
In total, we get a simulation cell of 1.65$\times$1.65$\times$25$\,$nm$^3$ with 2232 atoms.

\subsubsection{Si-MOS}

We generate the amorphous interface structures by \textit{ab-initio} molecular dynamics (AIMD) simulations, following our previous approach~\cite{Cvitko2023}. 
The thickness of the Si and SiO$_2$ layers were chosen such that the decay of the electron wavefunction into the semiconductor (along $z$) and into the oxide (along $-z$) makes the electron wavefunction negligible at the layer boundary (present in the model only) away from the interface.
A Si thickness of 10\,nm and an oxide thickness of 1\,nm proved sufficient to suppress boundary effects as the wavefunction decays quickly as a result of the large conduction band offset ($|\Psi|^2$ decays by about a factor of 10$^5$ per nm in SiO$_2$), in combination with the strong electric field.
We observe that after a roughly 0.5\,nm thick transition region at the crystalline Si, the amorphous SiO$_2$ shows geometric qualities (bond lengths and angles, densities) of bulk silicon dioxide~\cite{Cvitko2023}. 

To impose the periodicity in $z$-direction required by the DFT code, we adopt the repeated slab model: The top edge of the oxide layer and the bottom edge of the Si crystal are passivated with hydrogen and separated from their periodic image by a vacuum of 20\,nm. Subsequently, the whole simulation cell is relaxed in $x$ and $y$ direction with respect to the lattice parameter of bulk Si (our DFT setup gives a relaxed Si lattice parameter $a_0=0.544$\,nm which is very close to the experimental value of $a_0=0.543$\,nm).

\subsection{The hyperfine tensor}
\newcommand{\spinElectron}{\mathbf{S}}
\newcommand{\spinNucleus}{\mathbf{I}}
\newcommand{\Atensor}{\mathbf{A}}
\newcommand{\Aiso}{A}
\newcommand{\Aaniso}{\Atensor_\mathrm{ani}}
\newcommand{\AeffOne}{A^\perp}
\newcommand{\AeffTotal}[1]{\langle \Delta#1 \rangle}
The hyperfine interaction between the electron spin, described by a vector of operators $\spinElectron$, and the nuclear spin, described by vector $\spinNucleus$, is parameterized by the hyperfine tensor $\Atensor$,
\begin{equation}
\label{eq:Hamiltonian}
    H=\spinElectron \cdot \Atensor \cdot \spinNucleus.
\end{equation}
The tensor is the sum of isotropic and anisotropic contributions~\cite{ChrisG1993}
\begin{equation}
\label{eq:A=iso+aniso}
    \Atensor=\Aiso \,\mathrm{id}_3+\mathbf{A}_\mathrm{ani},
\end{equation}
where $\mathrm{id}_3$ is a three by three identity matrix.
The anisotropic part results from dipole-dipole interactions and is typically negligible in an $s$-type band, such as the silicon conduction  band~\cite{Assali11}.
Resulting from the Fermi contact interaction, the isotropic part is proportional to the electron density $|\Psi(\mathbf{r})|^2$ at the atomic core positioned at $\mathbf{r}$,\footnote{Equation \eqref{eq:Aiso} is the non-relativistic limit, in which the Fermi contact interaction takes the form of a delta function. The delta function is smeared to a finite size by relativistic effects. This smearing is important to remove spurious divergencies and is adopted in our DFT code. In the text, we restrict ourselves to the non-relativistic limit for simplicity.}
\begin{equation}
\label{eq:Aiso}
    \Aiso = \frac{4\mu_0}{3} g_e \mu_\mathrm{B} g_N \mu_N |\Psi(\mathbf{r})|^2,
\end{equation}
with the prefactor given by the vacuum permeability $\mu_0$, the Bohr and nuclear magneton $\mu_\mathrm{e}$ and $\mu_\mathrm{N}$, and the electron and nuclear $g$-factor in vacuum $g_\mathrm{e}$ and $g_\mathrm{N}$. 
For energies within a crystal band, the electron density depends on wavefunction envelope $\Phi$ and the Bloch part $u$,
\begin{equation*}
\qquad \qquad |\Psi(\mathbf{r})|^2 = |\Phi(\mathbf{r})|^2 |u(\mathbf{r})|^2. 
\end{equation*}
With the Bloch part normalized to one over the crystal unit cell, the latter factor is dimensionless. It is usually denoted as $\eta \equiv |u(\mathbf{r})|^2$.
It quantifies how much the electron is pulled towards the atomic core compared to the rest of the unit cell. Quoted values for the silicon conduction band are within $\eta \approx 160$--$190$ \cite{schliemann_electron_2003,fang_recent_2023}.\footnote{
\label{footnote_eta}
For germanium, the authors of Ref.~\cite{wilson_electron_1964} "have made an experimental estimate based on observed nuclear spin relaxation times which indicates $\eta_\mathrm{Ge}$ is an order of magnitude larger than $\eta_\mathrm{Si}$." The authors of Ref.~\cite{kerckhoff_magnetic_2021} say "We arrive at a value of 570 by an informal fit to our aggregated data across multiple devices with multiple quantum wells, and assume this value throughout, although we acknowledge that the present data leave this number about 30\% uncertain."
} 
An important difference between our work and Ref.~\cite{witzel_nuclear_2012} is that our DFT calculation delivers wave functions estimates reliable also near atomic cores. In Ref.~\cite{witzel_nuclear_2012}, the value of $\eta$ is an input that supplements the tight-binding method and the authors adopted the value from Ref.~\cite{wilson_electron_1964}.

We would like to stress that the separation of the envelope and Bloch parts is not possible in the barrier of Si-MOS. There is no meaningful definition of the Bloch wavefunction in the band gap of the amorphous oxide. Therefore, we do not express the hyperfine tensor as a function of $\eta$ below. In judging the `strength' of the hyperfine interaction, for example for oxygen versus silicon atoms, one can compare only the dimensionful quantity $\Atensor$. The effects of the decay into the barrier and the structure of the wavefunction on the atomic scale are inextricable.

\subsection{Decoherence in the ergodic limit}

We now estimate the dephasing of the electron spin coupled to a collection of nuclear spins through the hyperfine interaction given in Eq.~\eqref{eq:Hamiltonian}. This thermal bath induces fluctuations of the electron spin energy. The dephasing time $T_2^*$ is given by the variance of these fluctuations, $\left\langle \delta E^2 \right\rangle$, by a textbook formula~\cite{Abragam1989, burkard_semiconductor_2023}
\begin{equation}
\label{eq:T2star_TB}
    T_2^*=\hbar \frac{\sqrt{2}}{\sqrt{\left\langle \delta E^2 \right\rangle}}.
\end{equation}
Assuming the nuclear spins are unpolarized and thermally fluctuating, a standard derivation\footnote{The result in Eq.~\eqref{eq:fluctuations} relies on the assumption that the nuclear spins are `ergodic' in the nomenclature of Ref.~\cite{delbecq_quantum_2016}. It applies to experiments where the total time for the data collection, for example, to evaluate the dephasing time, is much larger than the auto-correlation time of the nuclear ensemble. The latter depends on the diffusion constant, which we estimate for the relevant atomic nuclei in App.~\ref{app:diffusion}.} \cite{merkulov_electron_2002} gives
the hyperfine-induced variance of the electron spin energy as  
\begin{equation}
\label{eq:fluctuations}
    \langle \delta E^2 \rangle = \sum_{i \in \mathrm{isotopes}} p_i \frac{I_i(I_i+1)}{3} \sum_{n \in \mathrm{isotope\,}i} |\mathbf{m} \cdot \Atensor_{i,n}|^2.
\end{equation}
Here the index\footnote{The isotope and atom indexes apply for the hyperfine tensor in Eqs.~\eqref{eq:Hamiltonian}-\eqref{eq:Aiso}. For example, we could reinstate these indexes on the right-hand side of Eq.~\eqref{eq:Aiso} by recognizing the isotope-dependence of $g_N \to g_{N,i}$ and the atom-dependence by $\mathbf{r} \to \mathbf{r}_{i,n}$. We have omitted indexes in Eq.~\eqref{eq:Hamiltonian}-\eqref{eq:Aiso} to simplify the notation.} $i \in \{^{73}\mathrm{Ge}, ^{29}\mathrm{Si} \}$ or $i \in \{^{17}\mathrm{O}, ^{29}\mathrm{Si} \}$ runs over spinful isotopes and $n$ runs over atoms of a given isotope. All atoms of a given isotope have the same nuclear spin magnitude $I_i$, being 9/2 for $^{73}$Ge, 5/2 for $^{17}$O, and 1/2 for $^{29}$Si, and the isotopic concentration $p_i$, being 7.8\% for $^{73}$Ge, 377 ppm for $^{17}$O and 4.7\% for $^{29}$Si in natural silicon and 50--800 ppm in isotopically purified one. Finally, $\mathbf{m}$ is the unit vector along the direction of the magnetic field and $|\cdot|$ is the Euclidean norm of a vector. We take $\mathbf{m}$ along the $z$ axis but expect no relevant directional dependence of the dephasing since the hyperfine tensors are dominated by the isotropic part.\footnote{Concerning the dephasing from the oxygen atoms in Si-MOS, we expect that any anisotropy of the hyperfine tensor would be averaged out by the random orientation of the atomic bonds in the amorphous oxide.}

\section{Results}

After explaining our approach, we move to the DFT results. Let us first examine the electronic structure at the conduction band edge. 

\subsection{The lowest subband of a Si heterostructure 2DEG}

It is well known that the six-fold degeneracy of the conduction band minimum of bulk silicon is lifted by the heterostructure confinement and, for Si/SiGe, by the in-plane strain in the Si layer.~\cite{Zwanenburg2013, burkard_semiconductor_2023}.
A four-dimensional subspace is raised in energy, leaving two so-called valley states at the conduction band minimum, at crystal momenta $k_0 \approx \pm \frac{2 \pi}{a_0} 0.83$~\cite{Zwanenburg2013, burkard_semiconductor_2023}. 
These valley states are further split by the interface~\cite{Boykin04, Cvitkovich2024}, that is by the electric field pressing the electron onto the barrier.\footnote{For zero electric field, they would be split by the Si/SiGe quantum well interfaces.} 
Our DFT results are in line with this picture. We obtain a conduction band offset of about 0.2\,eV and 2.7\,eV for Si/SiGe and Si-MOS, respectively. The valley splitting is below 300\,µeV in Si/SiGe and in the range of 1-3\,meV for the oxide interface.\footnote{Typical experimental observations fall within, or slightly below, this interval~\cite{Gamble2016, Zwanenburg2013}.}

\begin{figure}[tbp]
	\centerline{\includegraphics[width=\linewidth]{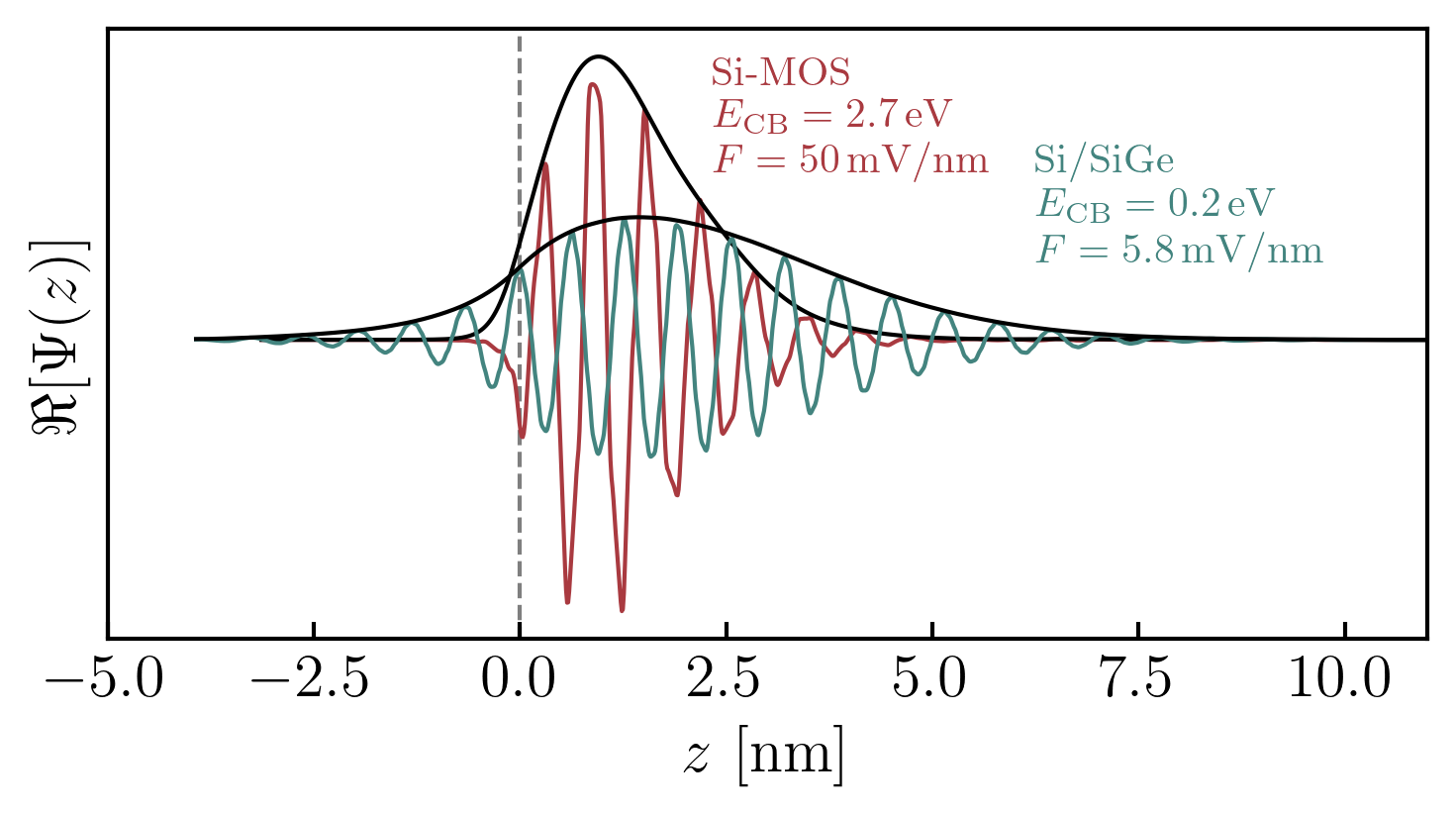}}
    \caption{
    Real-part of the cross-section averaged wavefunction at the conduction band minimum as obtained from DFT for the Si-MOS and Si/SiGe interface model, respectively. The first atom in the barrier is at $z=0$.
    When compared to Si-MOS, the Si/SiGe wavefunction extends further into the Si part of the simulation cell ($+z$) due to the lower electric field $F$. In addition, due to the lower barrier $E_{\mathrm{CB}}$, it also extends further into the SiGe barrier ($-z$).
    The associated envelopes $\Phi(z)$ (black lines) are obtained by low-pass filtering the wavefunction density with a Gaussian filter.
    }
    \label{fig:elec_struct}
\end{figure}

The ground state wavefunctions are plotted in Fig.~\ref{fig:elec_struct}. One can identify the oscillations assigned to the expected valley wave vector $k_0$.
The valley content of states at the conduction band minimum is also imprinted in the local density of states as shown in Fig.~\ref{fig:LDOS} in App.~\ref{app:dos} for an illustration.

\subsection{The magnitudes of the hyperfine couplings}

\begin{figure}[tbp]
	\centerline{\includegraphics[width=\linewidth]{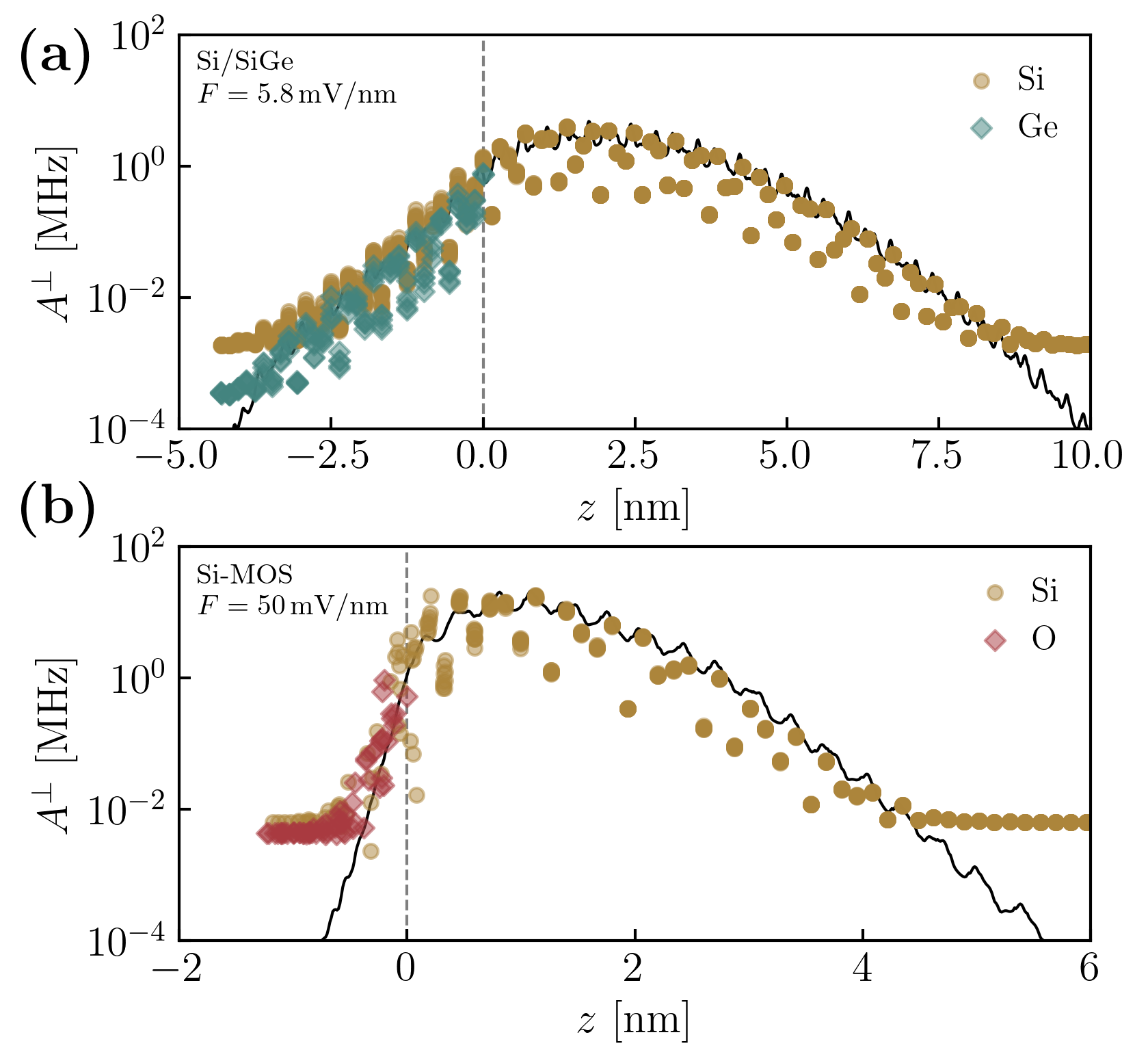}}
    \caption{
    Effective hyperfine interaction $A^\perp/2\pi \hbar$, for \textbf{(a)} Si/SiGe and \textbf{(b)} Si-MOS, of all atoms plotted along the $z$-direction of the simulation cell. The distribution can be largely attributed to the shape of the wavefunction density (not smoothed, black line). The boundary of the barrier layer interfacing the Si at $z=0$ is indicated by a dashed line. The interactions with barrier atoms further away from the interface subside quickly as a result of the exponential decay of the wavefunction in the barrier, while the saturation tail is due to anisotropic hyperfine interaction.
    }
    \label{fig:A_vs_z}
\end{figure}

Having the conduction-band wavefunction, including a reliable description of its shape near the atomic cores, we can evaluate the hyperfine tensors. For each atom, we obtain the tensor $\Atensor$. We define---loosely speaking---the `hyperfine strength', as the quantity entering the dephasing time formula\footnote{If one drops the anisotropic contributions, the `hyperfine strength' equals the hyperfine coupling $A$ given in Eq.~\eqref{eq:Aiso} and does not depend on the magnetic field direction $\mathbf{m}$.},
\begin{equation}
\label{eq:Aeff4}
    \AeffOne = |\mathbf{m} \cdot \Atensor|,
\end{equation}
where we again drop the indexes ${i,n}$.
Figure \ref{fig:A_vs_z} shows $\AeffOne$ plotted against the atoms' distance from the interface. It shows large variations which are easily explained by the proportionality of the dominant (isotropic) part to the electron wavefunction density given in Eq.~\eqref{eq:Aiso}. Thus, the overall trend in Fig.~\ref{fig:A_vs_z} reflects the confinement in $z$-direction. We attribute variations on top of this trend to Bloch oscillations and, especially relevant for the Si-MOS case, details of the atomic configuration at the interface. Though probably of little practical importance, we have also confirmed that the tails of the hyperfine strengths at the left and right boundaries are due to the anisotropic contribution. Here, as the electron wavefunction drops exponentially, the anisotropic hyperfine terms take over. Thus, the saturation at the wavefunction tails seems to be physical rather than a numerical artifact and it is reassuring that our numerical calculations can uncover these small tails.

The most important observation in Fig.~\ref{fig:A_vs_z} is that the hyperfine coupling to non-Si nuclei in the barrier is comparable to the coupling to silicon (if located at a similar distance from the interface). 
While we restrain from any quantitative fitting of $\eta$,\footnote{As already mentioned, $\eta$ is not well defined in the amorphous oxide. While well defined in crystalline semiconductors, fitting $\eta$ for both Ge and Si, is complicated by the interference of different valleys. We limit ourselves to a remark that we do not see any evidence of $\eta$ for Ge being much larger than for Si and, thus, our data give more support to the estimate for $\eta_\mathrm{Ge}/\eta_\mathrm{Si} \approx 3$ made in Ref.~\cite{kerckhoff_magnetic_2021} compared to $\eta_\mathrm{Ge}/\eta_\mathrm{Si} \approx 10$ made in Ref.~\cite{wilson_electron_1964} (see 
Footnote~\ref{footnote_eta}).} the lack of an essential difference between germanium/oxygen and silicon is obvious from the figure.
This is the crucial finding that decides the magnitude of the barrier-induced dephasing. 

To quantify their contributions, we assign the following extensive quantity to each isotope, as a short-hand notation for its contribution in Eq.~\eqref{eq:fluctuations},
\begin{equation}
\label{eq:Aeff}
    \AeffTotal{_i} \equiv \frac{I_i(I_i+1)}{3} \sum_{n \in \mathrm{isotope\,}i} |\mathbf{m} \cdot \Atensor_{i,n} |^2.
\end{equation}
Even though our numerical calculations and plots include the anisotropic contributions, we find that they are small. It is then useful to neglect them in the preceding equation, upon which we obtain
\begin{subequations}
\begin{align}
\label{eq:Aeff2}
    \AeffTotal{_i} &\approx \frac{I_i(I_i+1)}{3} \sum_{n \in \mathrm{isotope\,}i} \Aiso_{i,n}^2 \\
    &= c_i^2 \sum_{n \in \mathrm{isotope\,}i} |\Psi(\mathbf{r}_n)|^4, \label{eq:scaling}
\end{align}
where $c_i$ collects the constants from Eqs.~\eqref{eq:Aiso} and \eqref{eq:fluctuations},
\begin{equation}
\label{eq:Aeff3}
    c_i = \frac{4\mu_0}{9} I_i(I_i+1) g_e \mu_\mathrm{B} g_{N,i} \mu_N.
\end{equation}
\end{subequations}
The quantity $\AeffTotal{_i}$ gives the total sum of (squared) hyperfine couplings $\AeffOne$ for a given isotope, and, through that, the isotope contribution to the electron spin energy variance. 

With this notation, the dephasing time becomes,
\begin{equation}
\label{eq:T2star}
    T_2^*=\hbar \frac{\sqrt{2}}{\sqrt{\sum_i p_i \AeffTotal{_i}}}.
\end{equation}
One point is worth discussing here. In the DFT code, we do not distinguish among different isotopes of the same element. Correspondingly, in Fig.~\ref{fig:A_vs_z} we plot the hyperfine strengths for all atoms in the simulation cell. In reality, out of these, only a few atoms will be the spinful isotopes for each of the considered elements (Si, O, Ge). The quantity $p_i \AeffTotal{_i}$ is thus the average contribution to the electron energy variance from a given isotope, the average being over all possible distributions of the spinful isotopes within the set of atoms of a given element. We hint at this fact by using the angular brackets. By excluding the factor $p_i$, we have made the average quantity $\AeffTotal{_i}$ independent of the isotopic concentration, and thus useful when judging the effects of purification. The total hyperfine coupling $\AeffTotal{_i}$ is set by the heterostructure: the 2DEG width, the confinement field, and the barrier chemical composition.

With this clarification, let us look at the composition of $\AeffTotal{_i}$ for our structures. For Si-MOS, the maximal hyperfine coupling to an oxygen atom is two orders of magnitude smaller than the one for silicon. Nevertheless, as we will see below, the coupling is large enough to contribute to the dephasing appreciably. 
Finally, due to the exponential decay of the wave function in the barrier, interactions with atoms deeper than 1\,nm can be neglected. 

For Si/SiGe, the difference between the maximum hyperfine coupling of Si and Ge is less than one order of magnitude. Therefore, a stronger dephasing is expected from the Ge atoms in the barrier. In addition, the exponential decay in the barrier is significantly weaker compared to Si-MOS 
(a factor of about 10 per nm SiGe vs.~$10^5$ per nm oxide)
because of the much lower conduction band offset of only 200\,meV (vs.~2.7\,eV in Si-MOS). This difference results in appreciable interactions with Ge atoms located up to several nm inside the barrier.

\subsection{Rescaling the results for gated quantum dots}

\label{sec:scaling}

To estimate the dephasing time in a quantum dot spin qubit, we have to resolve one more issue. Namely, the size of a structure for which a DFT calculation is possible is limited. We focused on having a realistic description of the device along the $z$ direction, to faithfully describe the irregular interfaces. The lateral area of the modeled cell is thereby restricted to only about one or two square nanometers. However, gated quantum dot areas are hundred to thousand times larger.
We overcome this limitation by relying on the scaling of the hyperfine strengths with the electron wavefunction given in Eq.~\eqref{eq:scaling}. 
As already stated, and in line with claims in literature~\cite{Assali11}, we have found that the isotropic contribution dominates the total hyperfine strength $\AeffTotal{_i}$. Equation \eqref{eq:scaling} predicts that upon enlarging the lateral area of the simulation cell by a factor $N$, the quantity $\AeffTotal{_i}$ gets $N$-times smaller: while there will be (on average) $N$-times more atoms of a given isotope, each individual hyperfine strength drops by a factor of $N^2$, since the electron probability density is reduced as $|\Psi(\mathbf{r})|^2 \to |\Psi(\mathbf{r})|^2/N$.

With these considerations, we define an effective density of the total hyperfine coupling $\AeffTotal{^\mathrm{1d}}$ by multiplying the value obtained from our DFT code according to Eq.~\eqref{eq:Aeff} by the lateral area of the simulation cell. The isotope contribution to the quantum-dot electron energy variance is then
\begin{equation}
\label{eq:AeffTotal1D}
    \langle \delta E_i^2 \rangle = p_i \times \AeffTotal{_i^\mathrm{1d}} / S,
\end{equation}
where $S$ is the actual area of the quantum dot in the lateral dimensions. This approximation assumes that the structure of the interface does not change drastically upon moving along the lateral coordinates and that the quantity $\AeffTotal{_i}$ in a quantum dot is self-averaging. We expect that this is the case, with supportive evidence and further comments given in App.~\ref{app:scaling}.
 
\section{Discussion}
\label{lab:discussion}

\begin{table}[]
\caption{As a function of the electric field (top row), the middle three rows give the isotope contributions $\langle \Delta_i^\mathrm{1d} \rangle$ to the electron-spin energy variance in units of MHz$^2$nm$^2$. The bottom row gives the threshold isotopic concentration $p_\mathrm{^{29}Si}$ at which the barrier element (germanium or oxygen) contributes as much as silicon. These thresholds are calculated using Eq.~\eqref{eq:intersection} and natural isotopic concentrations, $p_\mathrm{^{73}Ge}=7.76$\% and $p_\mathrm{^{17}O}=377$ ppm.
}
\renewcommand{\arraystretch}{1.3}
\begin{tabular}{ c | @{\qquad}c @{\qquad}c @{\qquad}c @{\qquad}c @{\quad}c }
\toprule
   & \multicolumn{3}{c|}{Si/SiGe} & \multicolumn{2}{c}{Si-MOS} \\
\midrule
& \multicolumn{5}{c}{electric field $F$ [mV/nm]} \\
         &  4.5   & 5.8 & 7.5    & 43   &    50  \\
\midrule
isotope & \\
$^{29}$Si & 1185  & 1450 & 1563  & 3592 &  3766  \\
$^{73}$Ge & 13.9  & 64.5 & 163.3 &               \\
$^{17}$O  &       &      &       & 5.86 &  7.41  \\
\midrule
crossover [ppm]
          & 910   & 3453 & 5773 & 0.60  &  0.72  \\

\bottomrule
\end{tabular}
\label{tab:Delta}
\end{table}

Equation \eqref{eq:AeffTotal1D} allows us to extrapolate our DFT results to quantum dots of experimentally relevant sizes.
As an illustration, we consider a quantum dot with harmonic confinement with confinement lengths $l_x=l_y=30$\,nm, which corresponds to an effective QD area\footnote{
We use a standard definition of `effective quantum dot volume' $V$ given by $V^{-1} = \int |\Psi(\mathbf{r})|^4 \,\mathrm{d}\mathbf{r}$. Applying the definition within the $xy$ plane gives the effective area $S=\sqrt{2\pi}l_x\times\sqrt{2\pi}l_y$.} 
of $S\approx5500$\,nm$^2$.

\begin{figure}[tbp]
 \begin{overpic}[width=\linewidth]{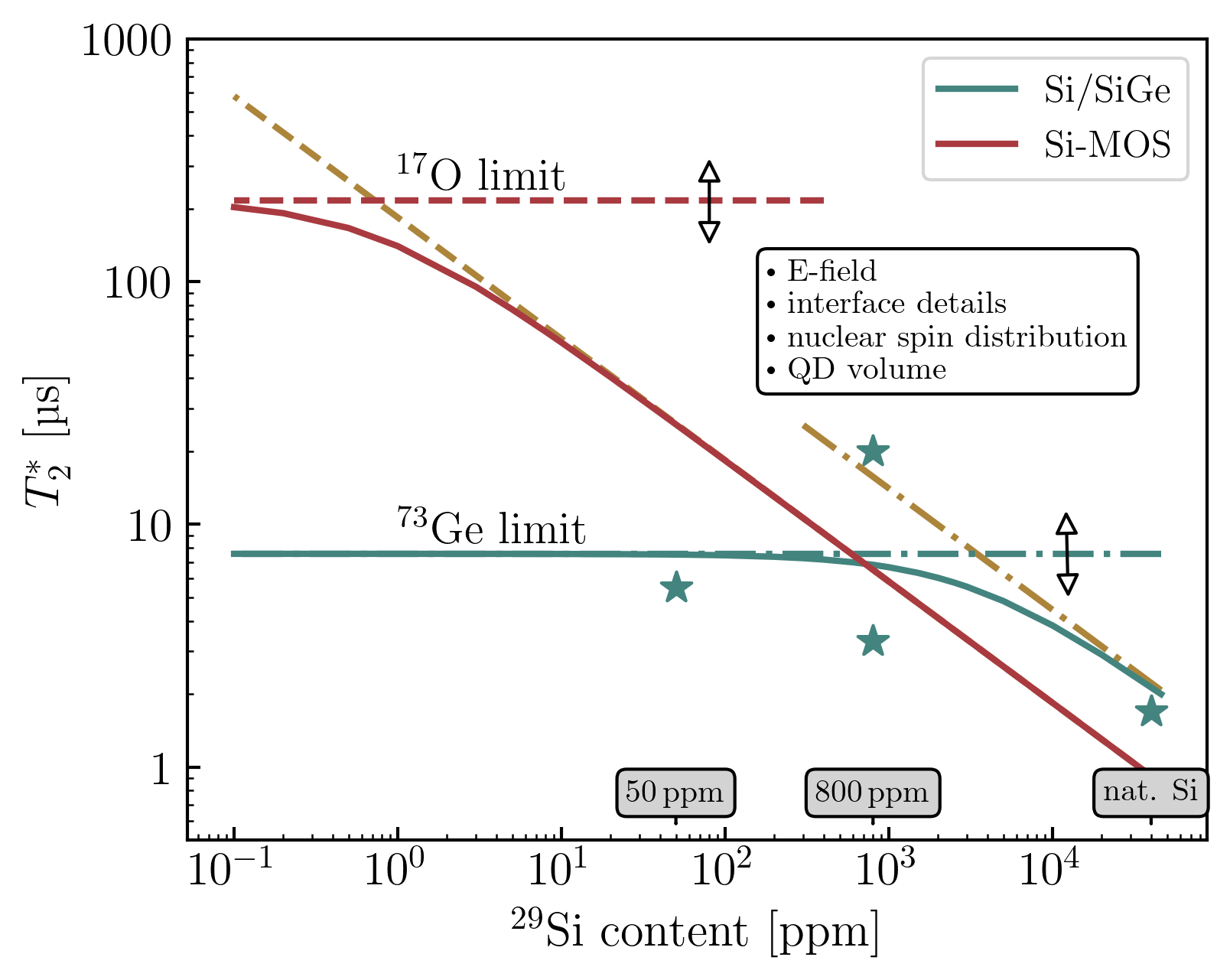}
    \put(73,42){\cite{yoneda_quantum-dot_2017,  struck_low-frequency_2020, Xue2022}}
    \put(73,24){\cite{Zhao2019}}
    \put(92,26){\cite{Takeda2021}}
    \put(57,30){\cite{tanttu_consistency_2023}}
\end{overpic}
    \caption{Dephasing times $T_2^*$ as a function of the amount of spinful $^{29}$Si nuclei. The field is $F=5.8$ mV/nm for Si/SiGe and $F=50$ mV/nm for Si-MOS. The increase of $T_2^*$ upon silicon purification is eventually limited by the germanium or oxygen atoms in the barrier. The uncertainty in the vertical position of the horizontal lines is pictorially represented by the double-headed arrows. The sources of this uncertainty are listed in the box and are explained and discussed in the main text below Eq.~(11). The stars represent experimental values for Si/SiGe at 800\,ppm from Ref.~\cite{yoneda_quantum-dot_2017,  struck_low-frequency_2020, Xue2022, Zhao2019}, at 50\,ppm from Ref.~\cite{tanttu_consistency_2023}, and in natural Si from Ref.~\cite{Takeda2021}.
    }
    \label{fig:decoherence}
\end{figure}
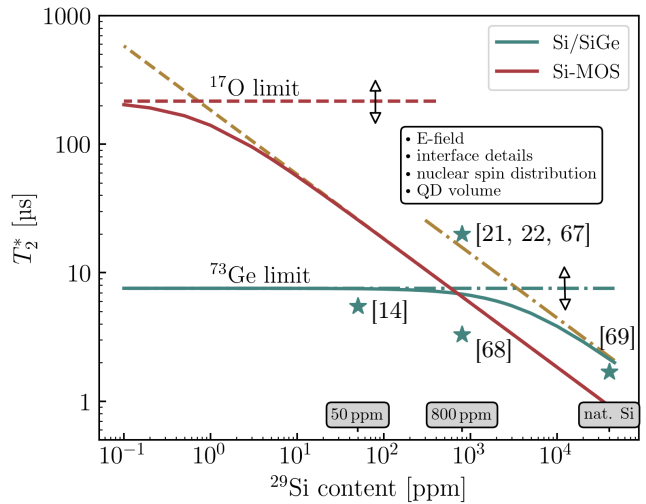

With this, we calculate the dephasing time $T_2^*$ according to Eqs.~\eqref{eq:T2star} and \eqref{eq:AeffTotal1D}.
We fix the isotopic concentration of Ge and O to their natural values, that is $p_{^{73}\mathrm{Ge}} = 7.76$\% and $p_{^{17}\mathrm{O}} = 377$\,ppm, and plot the electron spin dephasing time as a function of the $^{29}$Si content in Fig.~\ref{fig:decoherence}. 
We find a strong increase of $T_2^*$, boosting the coherence time by a factor of 20 when reducing the amount of $^{29}$Si from its natural abundance of 4.7\% down to the purest samples of nuclear-spin-free Si with 50\,ppm $^{29}$Si~\cite{tanttu_consistency_2023}.
The obtained $T_2^*$ of 1\,µs for natural Si and 20\,µs for purified Si at 50\,ppm are in line with earlier theoretical~\cite{Assali11, ChrisG1993} and experimental results~\cite{Muhonen2014,  Maurand2016, Struck2020, tanttu_consistency_2023, Tyryshkin2012, Veldhorst2014, Jock2018, Gamble2016}.

Further reduction of the $^{29}$Si content unveils the interactions with the barrier atoms. Identifiable as a crossover in Fig.~\ref{fig:decoherence}, they (germanium in Si/SiGe and oxygen in Si-MOS, respectively) will eventually dominate and limit the dephasing time achievable by silicon purification. The location of the crossover can be obtained by equating the contributions from the barrier atom and silicon using Eq.~\eqref{eq:AeffTotal1D},
\begin{equation}
\label{eq:intersection}
p_\mathrm{X} \AeffTotal{_\mathrm{X}^\mathrm{1d}}=
p_\mathrm{^{29}Si}  \AeffTotal{_\mathrm{^{29}Si}^\mathrm{1d}},
\end{equation}
where $\mathrm{X}$ stands for $\mathrm{^{73}Ge}$ or $\mathrm{^{17}O}$.

Solving for the silicon concentration using the DFT output gives results as listed in the last row of Tab.~\ref{tab:Delta}. In Si/SiO$_2$ we find $p_\mathrm{^{29}Si}$ around 1\,ppm, while in Si/SiGe the threshold is much higher, and in many devices it will be above 3000 ppm.


As already mentioned, these numbers are particular to the structure details. As is clear from Eq.~\eqref{eq:intersection}, the lateral size of the quantum dot is irrelevant for the crossover. However, the applied electric field is essential, especially for Si-MOS, as it determines how much the electron is pushed into the barrier. In our model the electric field has the same importance in the Si/SiGe variant even though for low electric fields, the penetration into the barrier will eventually be determined by the quantum well thickness. In both cases, increasing the electric field (or equivalently decreasing the quantum well thickness) enhances the barrier atoms' contribution, while the silicon contribution is affected less since the atoms inside the 2DEG dominate. As a result, the value of $p_\mathrm{^{29}Si}$ at the crossover increases. This behavior is on top of the expected dependence on the quantum dot area, which has been discussed around Eq.~\eqref{eq:AeffTotal1D}:  With other factors fixed, in a twice larger dot the individual hyperfine couplings $\AeffOne$ will be four times smaller, the total couplings $\AeffTotal{}$ twice smaller and the dephasing time twice longer. Such changes in quantum dot size would shift all curves plotted in Fig.~\ref{fig:decoherence} in the same way and do not change the horizontal position of the intersection.

We represent the dependence of the barrier atoms' contribution on various factors pictorially in Fig.~\ref{fig:decoherence} with the double arrow and list the factors in the text box. As discussed, the electric field is crucial for the barrier atoms' contribution as it determines the strength of the wave function tails. The details of the atomic arrangement at the interface are of similar relevance in Si/SiO$_2$. Although not explicitly investigated, we suspect
that analogous details might play a role in crystalline Si/SiGe, namely, how abruptly the Ge density changes (the interface width). We list this factor as `interface details' in the box of Fig.~\ref{fig:decoherence}. The next source of variations is the random location of the spinful isotopes, as we have already mentioned. This effect increases as the average number of spinful isotopes in the QD decreases and becomes pronounced when this number becomes of order one. Therefore, at the crossover, this effect will be strong in Si-MOS.
\footnote{We define the average effective number $n_\mathrm{X}$ of spinful atoms of isotope X contributing to the dephasing in a QD  using the inverse participation ratio. Namely, we collect the values of individual atoms contributions $\AeffOne_n$, normalize these numbers as a probability distribution, $\AeffOne_n \to P_n$ so that $\sum_n P_n =1$, and define $n_X = p_X / \sum_n P_n^2$. In this way, for an area $S=5500$ nm$^2$ and at the crossover, we obtain $n_\mathrm{^{29}Si}\approx3136$ and $n_\mathrm{^{73}Ge}\approx3661$ in Si/SiGe and $n_\mathrm{^{29}Si}\approx0.3$ and $n_\mathrm{^{17}O}\approx14.2$ in Si-MOS. 
}

Having all these error sources in mind, we anticipate a sizable spread of decoherence times in individual devices, even if they are dominated by nuclear noise. Comparison to experiments is further complicated by the fact that at low isotopic concentrations the coherence might be limited by other sources, for example charge noise~\cite{Yoneda2023, rojas-arias_spatial_2023}. We have included a few experimentally measured dephasing times in Fig.~\ref{fig:decoherence} for illustration. 

For Si purification above a few hundred ppm, Fig.~\ref{fig:decoherence} implies stronger dephasing in Si-MOS compared to Si/SiGe. The difference is not related to the barrier type but solely due to the stronger electric field. Pushing the QD against the interface generates a large spin density in the Si layers close to the interface, which translates into a stronger total hyperfine coupling.

\begin{figure}[tbp]
	\centerline{\includegraphics[width=\linewidth]{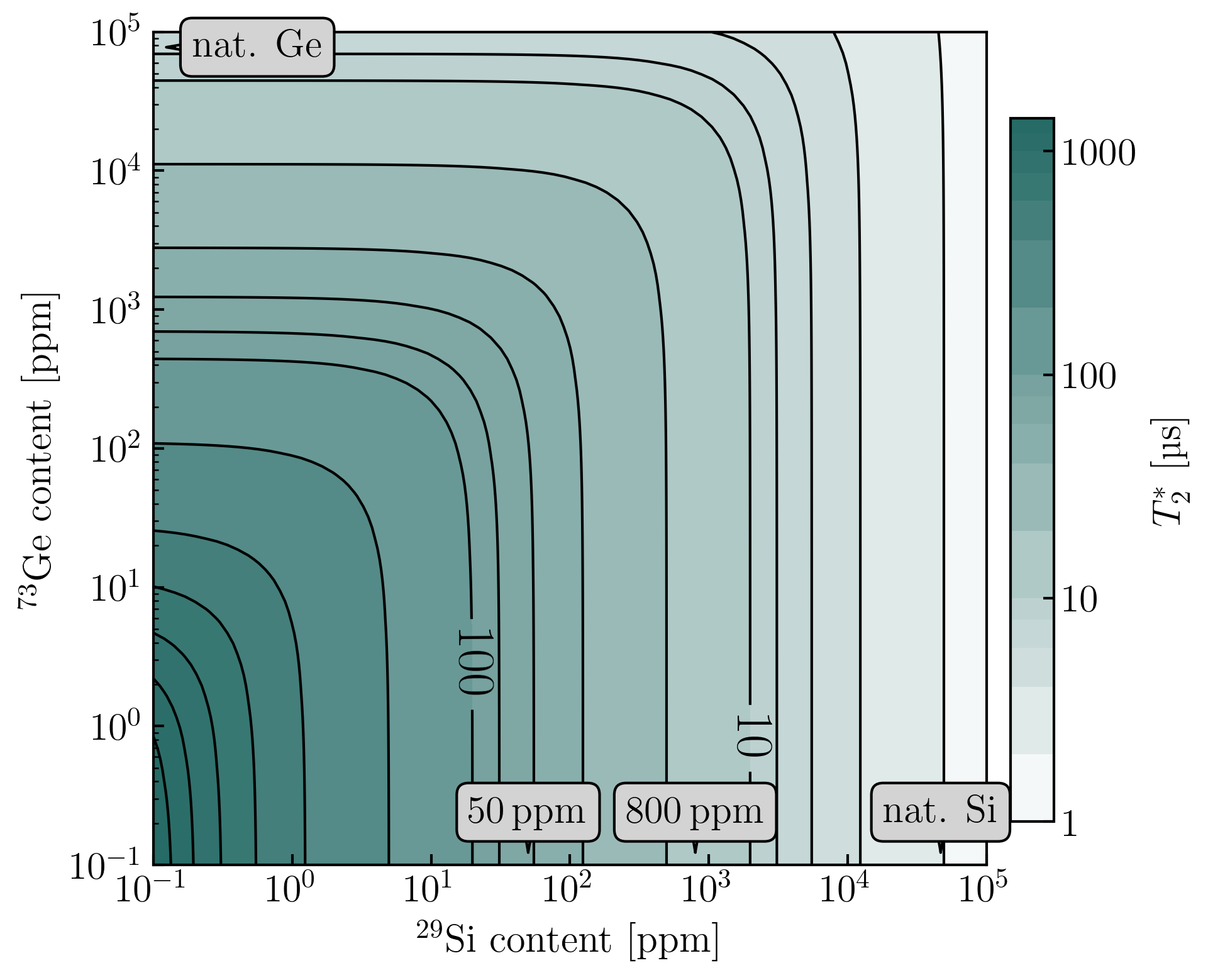}}
    \caption{Coherence times for Si/SiGe as a function of isotopic purification of $^{29}$Si and $^{73}$Ge. We plot $T_2^*$ evaluated according to Eq.~\eqref{eq:T2star} using the parameters from Tab.~\ref{tab:Delta} at $F=5.8$ mV/nm. 
    }
    \label{fig:Si_Ge_content}
\end{figure}

Finally, we discuss the isotopic purification of Ge, as its contribution to electron dephasing in Si/SiGe structures has been anticipated~\cite{witzel_nuclear_2012,Watzinger2018,struck_low-frequency_2020,kerckhoff_magnetic_2021}. Motivated by this prospect \cite{sigillito_electron_2015}, in Fig.~\ref{fig:Si_Ge_content} we plot the dephasing as a function of both Si and Ge isotopic content. The figure visualizes the crossover map for various isotopic concentrations. 

\section{Conclusions}
We calculate hyperfine interactions within a state-of-the-art \textit{ab-initio} framework and estimate the resulting dephasing of a spin qubit in Si-MOS and Si/SiGe quantum dots. The simulations include planar heterostructures with disordered interfaces at which a 2DEG is induced by an external electric field. We extract the hyperfine tensors for the interaction between the conduction-band electron and nuclear spins of every atom in the structure and examine the impact of isotopic purification on the coherence. We find that the improvement of the electron spin coherence time by isotopic purification is limited by the presence of spinful atoms in the barrier layer of the heterostructure once the $^{29}$Si content drops below a threshold value. The threshold strongly depends on the interface electric field. 
For the Si-MOS case, this threshold is around $1$\,ppm at $F=50$ mV/nm, below which the coherence-time is limited to 200\,µs by oxygen atoms.
In Si/SiGe with the natural abundance of 7.7\% $^{73}$Ge, the threshold at $F=5.8$ mV/nm is 3500\,ppm of $^{29}$Si, below which the coherence time is limited to 10\,µs by germanium atoms. 

\section{Acknowledgments}

This project has received funding from the European Research Council (ERC) under grant agreement no. 101055379. The computational results presented have been achieved using the Vienna Scientific Cluster (VSC).
L.C. gratefully acknowledges support from Institut français d'Autriche.
P.S. and D.L. acknowledge the support from CREST JST (JPMJCR1675) and D.L. from the Swiss National Science Foundation and NCCR SPIN grant No. 51NF40-180604.

\appendix

\section{The 2DEG subbands and the valley splitting}
\label{app:dos}

\begin{figure}[tbp]
	\centerline{\includegraphics[width=\linewidth]{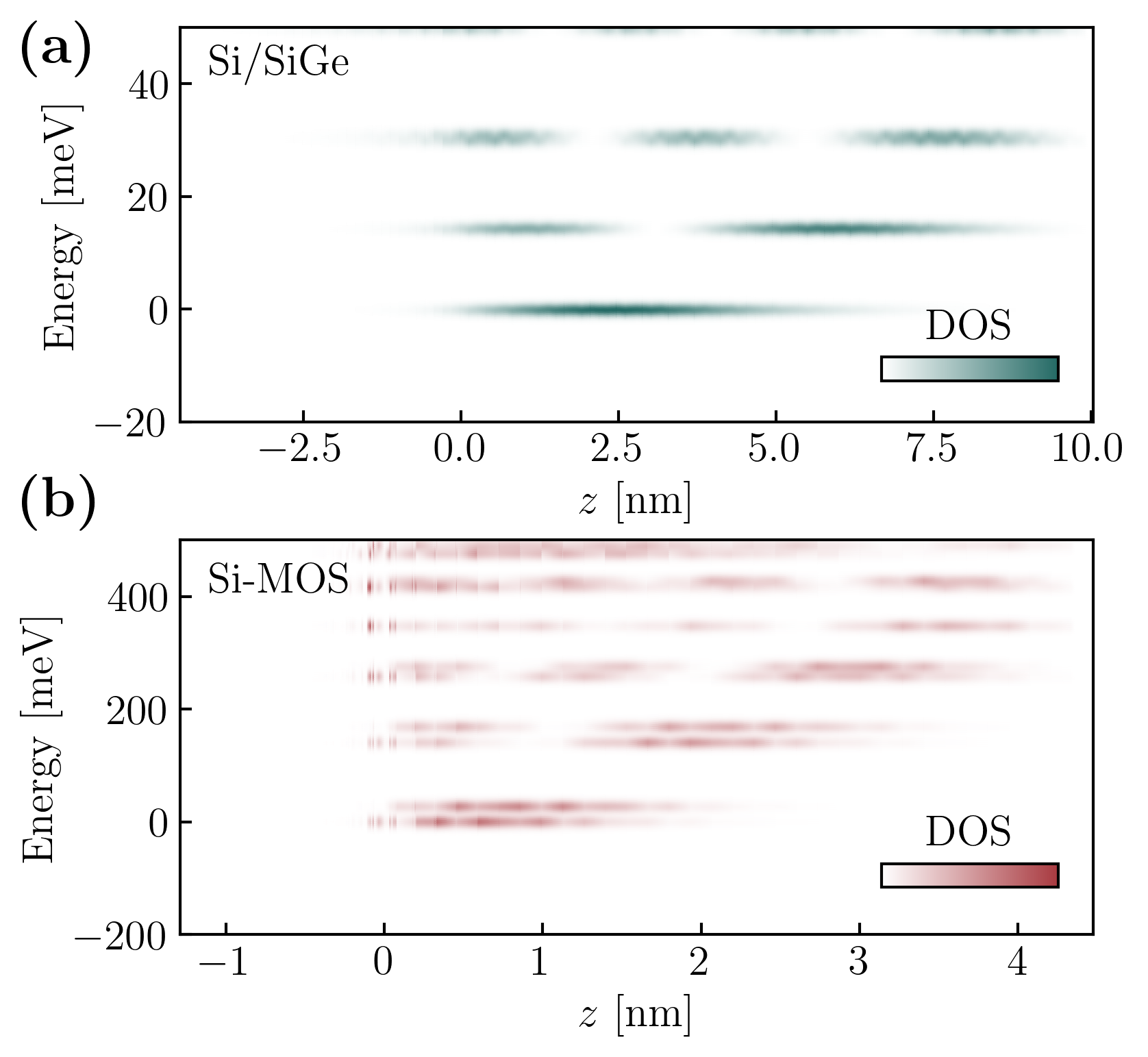}}
    \caption{
    Local density of states in the energy range of the conduction band minimum as obtained from DFT for \textbf{(a)} Si-MOS and \textbf{(b)} Si/SiGe. The interface is at $z=0$. The energy of the lowest state is set to zero. Each orbital state is comprised out of two valley states ($\pm z$) which are degenerate in bulk Si but couple here due to the presence of the interface.
    \textbf{(a)} For $F=50$\,mV/nm, the valley splitting in Si-MOS reaches values in the range between 1 and 10\,meV, depending on details of the interface~\cite{Cvitkovich2024}.
    \textbf{(b)} In Si/SiGe, the coupling is much weaker, with splitting below 300\,µeV, which is not discernible given the figure energy-axis scale. 
    }
    \label{fig:LDOS}
\end{figure}

For illustration, in Fig.~\ref{fig:LDOS} we show the local density of states at the conduction band minimum resolved in $z$ direction.
The low lying states come in pairs, each pair corresponding to a subband of the 2DEG. They arise as the eigenstates of the one-dimensional, approximately triangular, confinement of the interface: The wavefunction of the $n$-th excited state has $n$ nodes, if we count the ground state as the 0-th state. The subband energy splitting is determined by the electric field and is found to be 15\,meV for Si/SiGe and 150\,meV for Si-MOS. Each pair is further split by a much smaller energy (the valley splitting) and displays $k_0$ oscillations originating in the valley character of the state at the Si conduction band minimum. These oscillations are shifted by half a period in the two states of a pair.

\section{Scaling of the total hyperfine coupling with the dot area}
\label{app:scaling}

\begin{figure}[tbp]
	\centerline{\includegraphics[width=\linewidth]{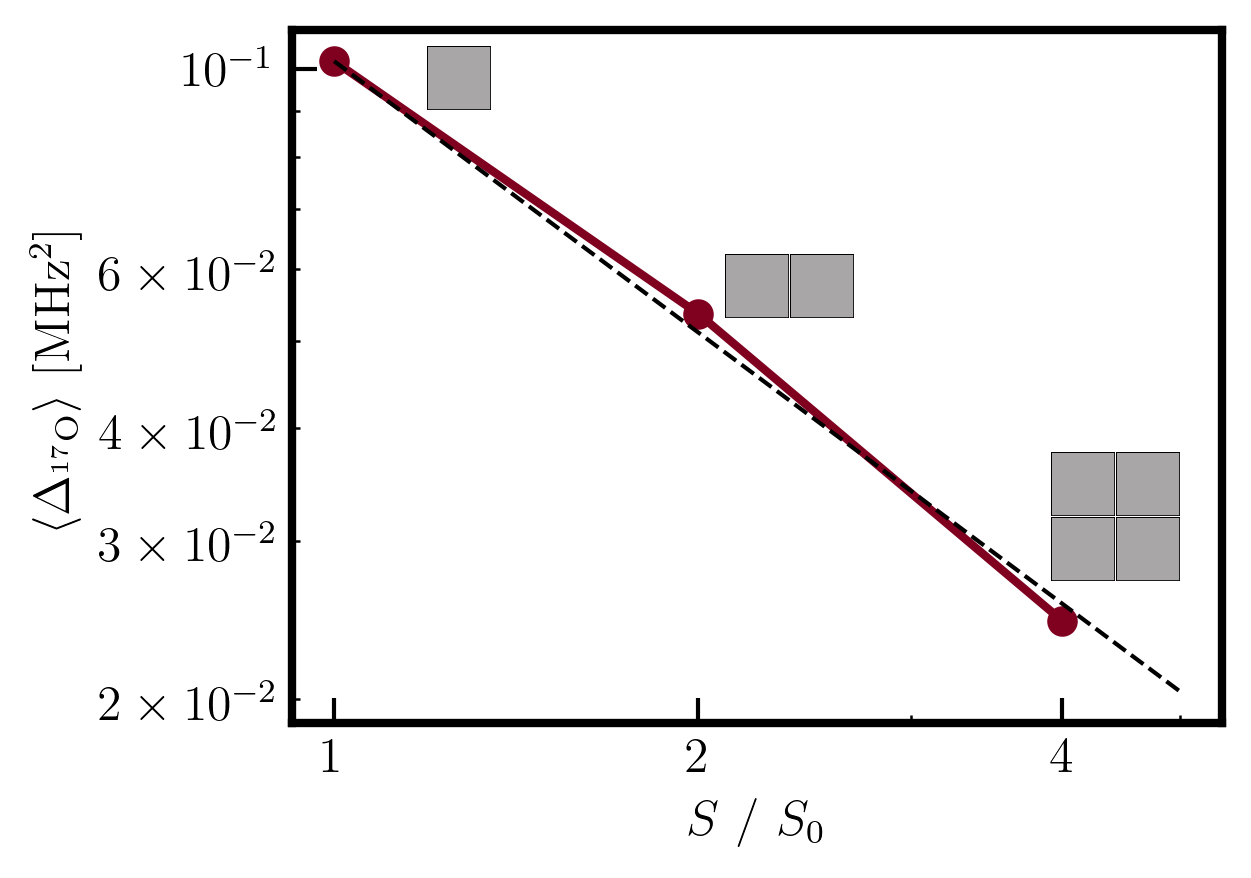}}
    \caption{Scaling of the simulation cell size. The total hyperfine strength for oxygen $\langle \Delta_\mathrm{^{17}O} \rangle$ in three runs with, respectively, single, double, and quadruple repetition of a unit block of 1.15\,nm$\times$1.15\,nm area. The dashed line is the expected scaling $\AeffTotal{} \propto 1/S$.}  
    \label{fig:scaling}
\end{figure}

The scaling of the total hyperfine strength $\AeffTotal{}$ with the quantum-dot area $S$, is illustrated in Fig.~\ref{fig:scaling}. It shows the results of three runs in which the original cell is replicated two and four times laterally. We observe the scaling predicted by Eq.~\eqref{eq:scaling} to good accuracy. Small deviations can be attributed to anisotropic contributions and numerical noise.

On the other hand, these results are not much more than a code sanity check since in those three runs we use the same configuration of atoms in each of the elementary squares depicted as the point labels in the figure. The inverse scaling with the quantum dot area (or volume) then follows from Eq.~\eqref{eq:scaling} and from the fact that the isotropic hyperfine coupling dominates, which is obvious from Fig.~\ref{fig:A_vs_z}. To check that the definition of $\AeffTotal{^\mathrm{1d}}$ is of practical use, we need to estimate the variance of $\AeffTotal{}$ with respect to atomic configurations. To this end, we have repeated the full simulation, including the molecular dynamics, a few times. The configurations of the amorphous oxide vary, resulting in a spread of $\AeffTotal{}$. The obtained values, given in Tab.~\ref{tab:four-runs-MOS}, can be considered independent samples of $\AeffTotal{}$ for our standard computation cell with 1.15\,nm lateral size. 

Apart from the variations in the atomic configurations in the oxide, the spread is further increased by varying hydrogen content at the interface. Paralleling their role in experiments, hydrogen atoms are used in DFT codes to passivate defects that occur as the structure is generated by \textit{ab initio} molecular dynamics. Namely, imposing cell-size restrictions together with periodic boundary conditions generates simulated structures with defects at the interface with a large probability. 
Typically, these defects are Si dangling bonds, which are bonds at which oxygen would sit in a defect-free SiO$_2$.
Due to the small cell area in our simulations, with less than five oxygen atoms at the interface on average, the structures with even a single defect show a significantly lower oxygen and an unrealistically high hydrogen concentration.\footnote{One defect, being one passivating hydrogen atom in our simulation cell area of about one nm$^2$, means a defect density about a hundred times more than what is expected from experiments~\cite{Stesmans1998}.}
Despite all these limitations, we conjecture from the four runs for Si-MOS listed in Tab.~\ref{tab:four-runs-MOS} that, first, the presence of hydrogen-passivated defects has a strong influence, and, second, defect-free structures (probably preferred in experiments) will have somewhat higher $\AeffTotal{}$ than the average calculated from also including structures with defects. The analogous data for Si/SiGe, given in Tab.~\ref{tab:four-runs-SiGe}, are in line with this guess. Here, the structure is crystalline and defect-free, and four runs gave values with a much smaller spread. The spread is due to variations in the placement of germanium atoms within the silicon matrix. Finally, the spread in the total effective hyperfine strength for silicon, in both Si/SiGe and Si-MOS, is well self-averaging even in our small simulation cell, as it is dominated by the contribution from the atoms in the defect-free crystalline silicon matrix of the 2DEG.

\begin{table}
\caption{Statistical uncertainty of $\AeffTotal{}$ for Si-MOS. The total hyperfine strength $\AeffTotal{}$ for $^{17}$O and $^{29}$Si as obtained in four full simulations at $F=50$\,mV/nm, including molecular dynamics building anew the semiconductor-oxide interface and the barrier oxide atomic structures.}
\begin{tabular}{@{\quad}c@{\quad}c@{\quad}c@{\quad}c@{}}
\toprule
& & \multicolumn{2}{c}{$\AeffTotal{} [\mathrm{MHz^2nm^2}]$}\\
\cmidrule{3-4}
run number&H atom count&$^{17}$O&$^{29}$Si\\
\midrule
\addlinespace[0.5em]
1   & 0 &  $7.41$  &   3766 \\
2   & 1 &  $3.43$  &   3736 \\
3   & 2 &  $2.7$   &   3912 \\
4   & 2 &  $2.8$   &   3852 \\
\midrule
 mean $\pm$ std&& $4.1 \pm 2.2$ & $3817\pm 80$ \\
\addlinespace[0.3em]
\bottomrule
\end{tabular}
\label{tab:four-runs-MOS}
\end{table}

\begin{table}
\caption{
Statistical uncertainty of $\AeffTotal{}$ for Si/SiGe. The total hyperfine strength $\AeffTotal{}$ for $^{73}$Ge and $^{29}$Si as obtained in four full simulations at $F=5.8$\,mV/nm.}
\begin{tabular}{@{\quad}c@{\quad}c@{\quad}c@{}}
\toprule
&\multicolumn{2}{c}{$\AeffTotal{} [\mathrm{MHz^2nm^2}]$}\\
\cmidrule{2-3}
run number&$^{73}$Ge&$^{29}$Si\\
\midrule
\addlinespace[0.5em]
1   &  $64.9$  &   1434 \\
2   &  $63.1$  &   1422 \\
3   &  $65.5$  &   1473 \\
4   &  $64.3$  &   1461 \\
\midrule
mean $\pm$ std& $64.5 \pm 1.0$ & $1448\pm24$\\
\addlinespace[0.3em]
\bottomrule
\end{tabular}
\label{tab:four-runs-SiGe}
\end{table}

\section{Estimation of the diffusion coefficient for $^{17}$O atoms}

\label{app:diffusion}

In the main text, we have estimated the effects of the oxygen and germanium nuclear spins on the quantum dot electron spin coherence assuming the so-called ergodic limit~\cite{delbecq_quantum_2016}. It applies if the total time over which data are collected (to evaluate the electron dephasing) is larger than the time scale over which the nuclear spin configuration changes due to its inherent dynamics. To make sure that the limit applies, the latter time scale must not be exceedingly long. One can understand this requirement also by pointing out that if the dynamics of the oxygen or germanium nuclear spins is very slow, the associated Overhauser field remains frozen during an experimental run probing the properties of the electron spin and does not contribute to its dephasing. 

To estimate the internal dynamics time scale, we assume that the nuclear spins undergo diffusion induced by their dipole-dipole interaction. The diffusion constant can be estimated by the method of moments~\cite{redfield_spatial_1959,slichter_principles_1996}. To this end, we evaluate formulas derived in Ref.~\cite{redfield_moment-method_1968}. We aim at an order-of-magnitude estimate, and adopt approximations accordingly.
\\

\subsection*{Expression for the diffusion constant derived by the method of moments}

\newcommand{\spinNucleusComponent}[2]{I^{#1}_{#2}}
\newcommand{\I}[2]{\spinNucleusComponent{#1}{#2}}
Previous work~\cite{redfield_moment-method_1968} considered an ensemble of alike nuclear spins in external magnetic field $B$ interacting pairwisely,\footnote{Our Eq.~\eqref{eq:B1} is Eq.~(19) of Ref.~\cite{redfield_moment-method_1968}, except that we define the Zeeman term with the opposite sign. In addition, we note that in experiments, a micromagnet inducing spatially dependent magnetic field is often employed to enable electrical manipulation of the electron spin. A large gradient could suppress the nuclear spin diffusion by inducing a mismatch of Zeeman energies for a spin pair that could otherwise make an energy-conserving spin flip-flop. Large inhomogeneous electric field gradients, coupling to the nuclear spins $I>1/2$ through the quadrupole interaction, or inhomogeneous Knight fields, could act similarly. Within our simple approach, we ignore these effects.}
\begin{equation}
\label{eq:B1}
H = -\hbar \gamma B \sum_i \I{i}{z} + \frac{1}{2}\sum_{i\neq j} \left( A_{ij}\I{i}{+}\I{j}{-} + B_{ij}\I{i}{z}\I{j}{z} \right),
\end{equation}
where for dipole-dipole interaction the couplings are
\begin{subequations}
\begin{align}
B_{ij} &= \frac{\mu_0}{4\pi} \gamma^2 \hbar^2 \frac{1-3\cos^2\theta_{ij}}{r_{ij}^3},\\
A_{ij} &= -\frac{1}{2} B_{ij},
\end{align}
\end{subequations}
expressed through quantities
\begin{subequations}
\begin{align}
\mathbf{r}_i & = \mathrm{position\, of\, nucleus\,} i,\\
\mathbf{r}_{ij} &= \mathbf{r}_{i} -\mathbf{r}_{j},\\
\cos^2\theta_{ij} & = \frac{|\mathbf{r}_{ij} \cdot \mathbf{z} |^2}{r_{ij}^2},
\end{align}
\end{subequations}
and with $\gamma = g \mu_N /\hbar$, $g$ being the nuclear $g$-factor, $\mu_N$ being the nuclear magneton, and $\mu_0$ being the vacuum magnetic permeability. We have also assumed that the magnetic field is applied along a unit vector $\mathbf{z}$ and define the nuclear spin operator in Cartesian components accordingly, with $\I{}{z} $$= \spinNucleus \cdot \mathbf{z}$.

Starting from Eq.~\eqref{eq:B1}, Ref.~\cite{redfield_moment-method_1968} derives a simplified expression for the diagonal elements of the diffusion tensor,\footnote{It is given in Ref.~\cite{redfield_moment-method_1968} as Eq.~(35), considering the relation $A_{ij}=-B_{ij}/2$.}
\begin{equation}
\label{eq:D-simplified}
D_{\mu \mu} \simeq \frac{\sqrt{\pi I(I+1)}}{4\sqrt{12}\hbar}\frac{\sum_j |\mathbf{r}_{ij} \cdot \boldsymbol{\mu}|^2 B_{ij}^2}{ \sqrt{\sum_k (B_{ik}-B_{jk})^2}}.
\end{equation}
Here in the upper (lower) sum the term $j=i$ is excluded (the terms $k=i$ and $k=j$ are excluded) and the index $\mu$ denotes a Cartesian component along a unit vector $\boldsymbol{\mu}$. Following Ref.~\cite{redfield_moment-method_1968}, we drop the cross term $B_{ik} B_{jk}$ from the sum in the denominator, which gives $\sum_k (B_{ik}-B_{jk})^2 \approx 2\sum_k B_{ik}^2$ upon renaming the summation index. With this change, we now further simplify the expression in Eq.~\eqref{eq:D-simplified}.

\begin{widetext}
\newcommand{\myslantfrac}[2]{${}^{#1}{\mskip -5mu/\mskip -3mu}_{#2}$}
\renewcommand{\arraystretch}{1.5}
\begin{table*}
\caption{
Estimated diffusion constants. The last column gives the diffusion constant calculated according to Eq.~\eqref{eq:D-final} for the isotope and material given in the first two columns. The table lists other quantities that enter the evaluated formula: the isotope nuclear spin $g$-factor $g$, spin magnitude $I$, (for crystalline materials) the number of atoms per unit cell with volume $a_0^3$, and the concentration (of the isotope among all isotopes of the given chemical element) $p$. The quantity $\rho_\mathrm{spin}$ is the isotope volume density, given by $\rho_\mathrm{spin} = p \rho_\mathrm{atom}$, and it defines the effective radius $r_\rho$ by Eq.~\eqref{eq:r_rho}.}
\begin{tabular}{@{\qquad}c@{\qquad}c@{\qquad}c@{\qquad}c@{\qquad}c@{\qquad}c@{\qquad}c@{\qquad}c@{\qquad}c@{\quad}}
\toprule
\multicolumn{2}{c}{system} & $g_N$ & $I$ & $\rho_\mathrm{atom}$&$p$ &$\rho_\mathrm{spin}$ & $r_\rho$ &$D$\\
isotope& lattice & dimensionless & dimensionless & $[\frac{1}{a_0^3}]$ & [ppm or \%] &$[\frac{1}{\mathrm{nm}^3}]$ & [\AA] & $[ \frac{\mathrm{nm}^2}{\mathrm{s}}]$ \\
\midrule
$^{17}$O & SiO$_2$ & -1.89 & \myslantfrac{5}{2}& -- & 380 ppm & 0.017 & 24 &8.5\\
$^{29}$Si & SiO$_2$ & -0.555 & \myslantfrac{1}{2}& -- & 4.67 \% & 1.02 & 6.2 &0.85\\
\midrule
\addlinespace[0.5em]
$^{29}$Si & diamond & -0.555 & \myslantfrac{1}{2}& 8 & 4.67 \% & 2.3 & 4.7 &1.1\\
$^{29}$Si & diamond & -0.555 & \myslantfrac{1}{2}& 8 & 800 ppm & 0.04 & 18 &0.29\\
$^{29}$Si & diamond & -0.555 & \myslantfrac{1}{2}& 8 & 50 ppm& 0.0025 & 46 &0.11\\
\midrule
\addlinespace[0.5em]
$^{29}$Si & SiGe & -0.555 & \myslantfrac{1}{2}& 4 & 4.67 \% & 1.1 & 6.1 &0.86\\
$^{73}$Ge & SiGe & -0.878 & \myslantfrac{9}{2}& 4 & 7.76 \% & 1.8 & 5.1 &15\\
\midrule
\addlinespace[0.3em]
$^{75}$As & GaAs & 1.44 & \myslantfrac{3}{2}& 4 & 100 \%& 22 & 2.2 &36\\
$^{71}$Ga & GaAs & 2.56 & \myslantfrac{3}{2}& 4 & 40 \%& 8.8 & 3.0 &83\\
$^{69}$Ga & GaAs & 2.02 & \myslantfrac{3}{2}& 4 & 60 \%& 13 & 2.6 &59\\
\bottomrule
\end{tabular}
\label{tab:D-final}
\end{table*}
\end{widetext}

First, we average over the direction of the diffusion, that is, the components of the diffusion tensor,
\begin{equation}
\label{eq:average1}
\langle \cdots \rangle_{\boldsymbol{\mu}} = \frac{1}{3} \sum_{\boldsymbol{\mu}\in \{\mathbf{x},\mathbf{y},\mathbf{z} \}} \cdots .
\end{equation}
This averaging simplifies the expression in the numerator of Eq.~\eqref{eq:D-simplified},
\begin{equation}
\label{eq:average1b}
\langle |\mathbf{r}_{ij} \cdot \boldsymbol{\mu}|^2 \rangle_{\boldsymbol{\mu}} = \frac{1}{3} r_{ij}^2.
\end{equation}
Second, we average over the orientation of the magnetic field with respect to the crystal lattice,
\begin{equation}
\label{eq:average2}
\langle\cdots\rangle_{\boldsymbol{z}} = \frac{1}{4\pi} \int_0^{2\pi} \mathrm{d}\phi \int_{0}^{\pi} \sin\theta \, \mathrm{d}\theta \cdots .
\end{equation}
The average of the angular factors present in Eq.~\eqref{eq:D-simplified} is
\begin{equation}
\label{eq:average2b}
\langle (1-3\cos^2 \theta)^2 \rangle_{\boldsymbol{z}} = \frac{4}{5}.
\end{equation}
Applying Eq.~\eqref{eq:average1b} in the numerator, and Eq.~\eqref{eq:average2b} separately in the numerator and the square of the denominator, we simplify Eq.~\eqref{eq:D-simplified} into
\begin{equation}
D \simeq \frac{\sqrt{\pi I(I+1)}}{4\sqrt{24}\hbar} \frac{\mu_0}{4\pi} \gamma^2 \hbar^2 \frac{\sum_j (1/3) \times (4/5) r_{ij}^{-4}}{ \sqrt{\sum_j (4/5) \times r_{ij}^{-6}}}.
\end{equation} 
Next, we replace the discrete sums with integrals, assuming that the spins $j$ are distributed approximately uniformly in space with density $\rho_\mathrm{spin}$,
\begin{equation}
\label{eq:average3}
\biggl< \sum_j r_{ij}^{-n} \biggr>_\rho = \int_{r_\mathrm{min}}^\infty 4\pi r^2 \rho_\mathrm{spin} \times r^{-n} = 4\pi \rho_\mathrm{spin} \frac{r_\mathrm{min}^{3-n}}{n-3},
\end{equation}
for $n>3$. Implementing the procedures defined in Eqs.~\eqref{eq:average1}, \eqref{eq:average2}, \eqref{eq:average3}, the average of  Eq.~\eqref{eq:D-simplified} is
\begin{equation}
\langle D \rangle_{\boldsymbol{\mu},\mathbf{z},\rho} \simeq \frac{\pi}{3\sqrt{40}} \sqrt{I(I+1)}\frac{\mu_0}{4\pi} \gamma^2 \hbar  \sqrt{\rho_\mathrm{spin} r_\mathrm{min}}.
\end{equation}
We define the `nearest-neighbor distance' as 2$r_\rho$, twice the radius of a sphere that has a volume equal to the volume per single spinful isotope. The latter is defined by the relation
\begin{equation}
\label{eq:r_rho}
\rho_\mathrm{spin}^{-1} = 
\frac{4\pi}{3} r_\rho^3.
\end{equation}The density $\rho_\mathrm{spin}$ equals the density of a given atomic element in the lattice $\rho_\mathrm{atom}$ (for example, for Si it is 8 atoms per unit cell of the diamond lattice) times the isotopic ratio of a given isotope $p$ (for example, for $^{29}$Si with it would be 4.7\% for Si with natural isotope composition).

As the final step, we choose the short-distance cutoff $r_\mathrm{min} = r_\rho$. This choice means that in the integral in Eq.~\eqref{eq:average3} we implement the condition $j\neq i$ by excluding a sphere around spin $i$ with radius $r_\mathrm{\rho}$, excluding a spherical volume corresponding to a single nuclear spin in the lattice. We obtain 
\begin{equation}
\label{eq:D-final}
\langle D \rangle_{\boldsymbol{\mu},\mathbf{z},\rho} \simeq \frac{\sqrt{\pi}}{4\sqrt{30}} \sqrt{I(I+1)}\frac{\mu_0}{4\pi} \gamma^2 \hbar  \frac{1}{r_\rho}.
\end{equation}

\subsection*{Numerical values for the diffusion coefficient in various materials}

We evaluate Eq.~\eqref{eq:D-final} for several scenarios in Tab.~\ref{tab:D-final}. While we are interested in the diffusion of nuclear spins of $^{17}$O and $^{73}$Ge, we include other standard spin-qubit materials and elements, namely Si, SiGe, and GaAs, to check the prediction of Eq.~\eqref{eq:D-final}. The table lists the quantities that enter the formula, and gives the averaged diffusion constants in the last column. They are of order ten(s) of $\mathrm{nm}^2/\mathrm{s}$ for GaAs isotopes and $^{73}$Ge in SiGe. The spins of silicon $^{29}$Si diffuse slower, about an order of magnitude, due to its lower spin magnitude, lower $g$-factor, and lower isotopic concentration. Surprisingly, the diffusion of $^{17}$O is not slow, despite its minuscule isotopic concentration. In SiO$_2$, it is an order of magnitude larger than that of Si. Again, the difference originates from the larger spin and $g$-factor of oxygen. We thus believe that concerning the expected time scales for the internal dynamics of Overhauser fields, there are no qualitative differences between oxygen in SiO$_2$ and isotopes of GaAs or SiGe. The latter fields have been observed to decorrelate on scales of seconds to minutes. We conclude that neither the oxygen- nor germanium-induced Overhauser field is frozen in a typical experiment and they both contribute to dephasing. 

As an illustration, we further quantify the diffusion of oxygen nuclear spins by converting the diffusion constant value given in Tab.~\ref{tab:D-final} to a correlation time and a power spectral density. We do this conversion only for the Si-MOS case. The presented quantitative characteristics should be taken as ballpark estimates. 

\subsection*{Autocorrelation time of the Overhauser field}

We start with the auto-correlator of the electron energy fluctuations due to the Overhauser field,
\begin{equation}
\label{eq:correlator}
\langle \delta E(t) \delta E(t+\tau)\rangle \equiv C(\tau) = \frac{p \langle \Delta \rangle }{\prod_{\mu} (1+\gamma_\mu |\tau|)^{1/2}}.
\end{equation}
We conjecture this equation from Eq.~(A5)  in Ref.~\cite{rojas-arias_spatial_2023}. We also amend it for our case and notation: The index $\mu \in \{x,y,z\}$ enumerates Cartesian coordinates, the quantity $\Delta$ is the oxygen-induced Overhauser field contribution defined in Eq.~\eqref{eq:Aeff} and $\gamma_\mu=2D_{\mu\mu}/l_\mu^2$. For $D_{\mu\mu}$ we take the value from Tab.~\ref{tab:D-final}. In Ref.~\cite{rojas-arias_spatial_2023}, the quantities $l_\mu$ denoted the quantum dot confinmenet lengths. Instead, here we take $l_\mu$ equal to the typical distance between spinful isotopes, $2r_\rho$.\footnote{Since the value of $2r_\rho$ is about 5 nm, the difference to taking quantum dot size would not be radical.} We stipulate that, since there are very few spinful oxygen atoms in a given quantum dot, as soon as the nuclear spin diffuses to the next available position in the oxide, it will not be coupled to the electron anymore, with a high probability. These considerations result in the correlation time scale
\begin{equation}
\tau_0 \equiv \frac{1}{\gamma} = \frac{2 r_\rho^2}{D},
\end{equation}
which evaluates to about 1 second for the parameters in the first line of Tab.~\ref{tab:D-final}. This is the expected time scale over which the $^{17}$O nuclear spin configuration, as seen by the electron spin, changes.

\begin{figure}[tbp]
\centerline{\includegraphics[width=\linewidth]{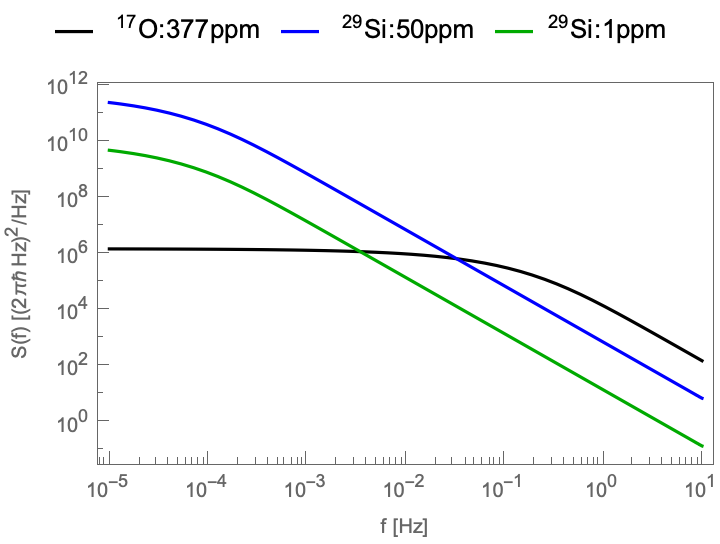}}
\caption{
\label{fig:psd}
Power spectral density.
The curves show the integral in Eq.~\eqref{eq:psd-integral} evaluated analytically for $n=3$, adopting the isotopic concentration $p$ as denoted in the curve labels, and the area $S=2\pi \times (30\,\mathrm{nm})^2$. Further, we took, for oxygen, $D=8.5$ nm$^2$/s and $l_\mu=4.8$ nm to evaluate $\tau_0$ and $\AeffTotal{^\mathrm{1d}}=4.1 \times (2\pi \hbar)^2$ MHz$^2$ nm$^2$ to evaluate $\AeffTotal{}$. For silicon, we took $D=0.29$ or $0.11$ nm$^2$/s (depending on $p_{^{29}\mathrm{Si}}$, as given in Tab.~\ref{tab:D-final}), $l_\mu=(4\times 2\pi \times 30^2)^{1/3}$ nm, and $\AeffTotal{^\mathrm{1d}}=3817 \times (2\pi \hbar)^2$ MHz$^2$ nm$^2$.
}
\end{figure}

\subsection*{Power spectral density}

A second quantity of immediate interest is the resulting noise spectral density. Defined by
\begin{equation}
P(\omega) \equiv \int_{-\infty}^\infty C(\tau) \exp( i \omega \tau) \, \mathrm{d}\tau,
\end{equation}
we get it in rescaled units $\tilde{\omega} = \omega \tau_0$ as
\begin{equation}
\label{eq:psd-integral}
P(\tilde{\omega}) = P_0  \int_{0}^\infty \frac{2 \cos( \tilde{\omega} x)}{\left(1+x \right)^{n/2}} \, \mathrm{d}x,
\end{equation}
where the scale is
\begin{equation}
\label{eq:P0}
P_0 = p \langle \Delta \rangle \tau_0.
\end{equation}
We have given the above equation for a variable number of dimensions $n$. The quantity $n$ is defined as the number of axes along which the diffusion can proceed, and enters our formula as the domain of the index $\mu$ in Eq.~\eqref{eq:correlator}. For example, if for some reason the diffusion is only along $z$, the index $\mu\in \{ z\}$ and $n=1$. The case $n=3$ being $\mu\in \{ x,y,z\}$ corresponds to the three-dimensional diffusion considered in the above.

Taking $\AeffTotal{_{^{17}\mathrm{O}}^{1\mathrm{d}}} = (2\pi \hbar  \times 4.1\, \mathrm{Mhz})^2 / \mathrm{nm}^2$ from Tab.~\ref{tab:four-runs-MOS} and $S=2\pi \times (30\, \mathrm{nm})^2$, we get $ P_0=3.7 \times 10^{5} (2\pi \hbar \times \mathrm{Hz})^2/\mathrm{Hz}$. The corresponding PSD is plotted in Fig.~\ref{fig:psd} using $n=3$. We compare it with the PSD due to $^{29}\mathrm{Si}$ for two concentrations. The 1\,ppm shows a PSD that, integrating over all frequencies, is approximately of the same strength as that of oxygen. Because of the difference in the characteristic times $\tau_0$, which is much larger for the diffusion of silicon nuclear spin, the shape of the curves is different. We also include the 50 ppm case. Interestingly, even though the total noise power is now much larger for silicon, the difference in the curve shapes still makes the oxygen contribution dominant at high enough frequencies. The reason is the same, the different characteristic times $\tau_0$. 
 

\bibliography{my.bib,2024-Cvitkovich.bib}

\end{document}